\documentclass[proceedings]{JHEP3} 
\PrHEP{ hep2001}

\usepackage{epsfig,multicol}            

\newbox\mybox
\newcommand\fverb{\setbox\mybox=\hbox\bgroup\verb}
\newcommand\fverbdo{\egroup\medskip\noindent\fbox{\unhbox\mybox}\ }
\newcommand\fverbit{\egroup\item[\fbox{\unhbox\mybox}]}

\newcommand{\nco}{\newcommand}
\nco{\beq}{\begin{equation}} \nco{\eeq}{\end{equation}}
\nco{\beqa}{\begin{eqnarray}} \nco{\eeqa}{\end{eqnarray}}
\nco{\lra}{\leftrightarrow}

\nco{\sss}{\scriptscriptstyle} \nco{\dphi}{\varphi}
\nco{\lsim}{\mbox{\raisebox{-.6ex}{~$\stackrel{<}{\sim}$~}}}
\nco{\gsim}{\mbox{\raisebox{-.6ex}{~$\stackrel{>}{\sim}$~}}}

\def\pref#1{(\ref{#1})}
\def\la{\lsim}
\def\ga{\gsim}

\def\vector#1{{\vec{#1}}}
\def\rot{{\rm curl}\,}
\def\div{{\rm div}\,}
\def\pref#1{(\ref{#1})}

\newcommand{\mnras}{MNRAS}
\newcommand{\aap}{A\&A}

\newcommand{\AHEP}{Instituto de F\'{\i}sica Corpuscular --
  C.S.I.C./Universitat de Val{\`e}ncia \\
  Edificio Institutos de Paterna, Apt 22085,
  E--46071 Val{\`e}ncia, Spain}
\newcommand{\IZMIRAN}{The Institute of Terrestrial Magnetism,
Ionosphere and 
Radio Wave Propagation \\ Russian Academy of Sciences,
  IZMIRAN, Troitsk, Moscow region, 142190, Russia}

\title{Neutrino Oscillations, Fluctuations and Solar Magneto-gravity Waves}

\author{\speaker{C.P.~Burgess},${}^1$ N. S. Dzhalilov,$^2$ M. Maltoni,$^3$
T.I. Rashba,$^{2,3}$ V. B. Semikoz,$^{2,3}$ M. Tortola,$^3$ and
J.~W.~F. Valle$^3$\\
        ${}^a$ Physics Department, McGill University,
        3600 University Street, Montr\'eal, Qu\'ebec, Canada H3A 2T8.\\
        ${}^2$ \IZMIRAN\\
        $^3$ \AHEP }

\conference{AHEP 2003, Valencia, Spain, October 2003.}

\abstract{This review has two parts. The first part summarizes the
current observational constraints on fluctuations in the solar
medium deep within the solar Radiative Zone, and shows how the
KamLAND and SNO-salt data combine to make the experimental
determination of the neutrino oscillation parameters largely
insensitive to prior assumptions about the nature of these
oscillations. As part of a search for plausible sources of solar
fluctuations to which neutrinos could be sensitive, the second
part of the talk summarizes a preliminary analysis of the
influence of magnetic fields on helioseismic waves. Using
simplifying assumptions which should apply to modes in the solar
radiative zone, we find a resonance between Alfv\'en waves and
helioseismic $g$-modes which potentially modifies the solar
density profile fairly significantly over comparatively short
distance scales, too narrow to be ruled out by present-day
analyses of $p$-wave helioseismic spectra. }

\begin{document}

\section{Introduction}

Over the past decade a consistent picture of neutrino mixing has
emerged, in which the oscillations responsible for both the solar
\cite{sol02,Ahmad:2002jz,Fukuda:2002pe} and atmospheric
\cite{atm02,Fukuda:1998mi} neutrino deficits are consistent with
the results of terrestrial neutrino-disappearance experiments
\cite{kamland,k2k}. Experiments are now moving beyond the
discovery phase and into a measurement phase during which the
oscillation parameters are being determined with unprecedented
precision.

Now that a coherent picture of neutrino properties seems to be in
place, solar neutrinos can be used in the way the early
investigators originally envisaged \cite{Bahcall}: as probes of
the deep solar interior. Traditionally, the solar neutrino flux
was believed to be exclusively sensitive to the properties of the
solar medium deep inside the core since this is the region where
the neutrinos are produced. The ${}^8B$ neutrino flux predicted by
solar models is very sensitive to the temperature in the solar
core and the extremely weak neutrino interactions ensure that
their observed energy spectrum is not affected by their passage
through the solar medium (in the absence of oscillations). Indeed,
we now know that the good agreement between solar-model
predictions and the measured flux of neutrinos of all three
species strongly constrains deviations of the core temperature
from solar-model predictions. Better yet, these models are
independently precisely tested by comparison with helioseismic
measurements, again with good agreement between the models and
observations
\cite{Castellani:1997pk,Christensen-Dalsgaard:2002ur}.

\subsection{Sensitivity to Solar Fluctuations}

As was realized in the mid-nineties
\cite{Balantekin:1996pp,Nunokawa:1996qu,Burgess:1996,Bamert:1997jj},
the existence of resonant neutrino oscillations potentially makes
the observed solar neutrino flux sensitive to other parts of the
solar environment. In particular, this flux can be sensitive to
fluctuations in the solar medium at the place where the neutrino
resonance occurs. As a result, some of the properties of the solar
medium at the neutrino resonance point may be inferred by
comparing the measured neutrino energy spectrum with what is
predicted by neutrino oscillations.

Of course such an inference of solar properties depends on having
precise terrestrial observations of neutrino oscillation
parameters, since these are required in order to cleanly predict
the neutrino energy spectrum. Conversely, in the absence of
terrestrial measurements, the precision of any determination of
neutrino oscillation parameters using solar neutrinos can be
degraded by the possible existence of solar fluctuations which
affect the observed neutrino signal \cite{us,them,usonSNO}.

It is the purpose of the first part of this review to show that
terrestrial \cite{kamland} and solar \cite{SNOsalt} neutrino
measurements have recently turned a corner inasmuch as they are
now sufficiently accurate to allow the removal of solar
uncertainties from the inference of neutrino oscillation
properties. Conversely, the comparison of solar neutrino data with
terrestrial observations now provides a clean window onto a new
part of the deep solar interior.

\subsection{Solar Magneto-gravity Waves}

Aficionados have not been too alarmed by the necessity to assume
the absence of solar fluctuations in order to infer neutrino
properties, for several reasons. First, helioseismic measurements
were known to constrain deviations of solar properties from
Standard Solar Model predictions at better than the percent level.
Second, the first studies of the implications for neutrino
oscillations of radiative-zone helioseismic waves
\cite{Bamert:1997jj} showed that they were very unlikely to have
observable effects. Third, no other known sources of fluctuations
seemed to have the properties required to influence neutrino
oscillations.

All three of these points have been re-examined in recent years
and although it may yet turn out that solar fluctuations do not
observably perturb neutrino oscillations, the possibilities are
more promising than had been presumed earlier. Direct helioseismic
bounds turn out to be insensitive to fluctuations whose size is as
small as those to which neutrinos are sensitive
\cite{Castellani:1997pk,Christensen-Dalsgaard:2002ur} (which turn
out to be those whose size is comparable to the neutrino
oscillation length in matter: several hundreds of km).
Furthermore, recent studies of magnetic fields deep inside the
solar radiative zone \cite{heliomag} have identified potential
fluctuations to which neutrinos might be sensitive after all (due
to a resonance between Alfv\'en waves and helioseismic $g$-modes).
It is the summary of this last observation which is the topic of
the second half of this review.

We now turn to a more detailed description of these two topics.


\section{Sensitivity to Solar Fluctuations}
The standard description of MSW oscillations
\cite{Wolfenstein:1977ue} amount to the use of a mean-field
approximation for the solar medium. The corrections to this
mean-field approximation are due to the fluctuations in the solar
medium about this mean, and the leading interaction of neutrinos
with these fluctuations are parameterized by the electron-density
autocorrelation, $\langle \delta n_e(t) \delta n_e(t') \rangle$,
measured along the neutrino trajectory
\cite{Balantekin:1996pp,Nunokawa:1996qu,Burgess:1996,Bamert:1997jj}.
 \begin{figure}[htb!]
  \begin{center}
  \def\epsfsize#1#2{0.35#1}
  \centerline{\epsfbox{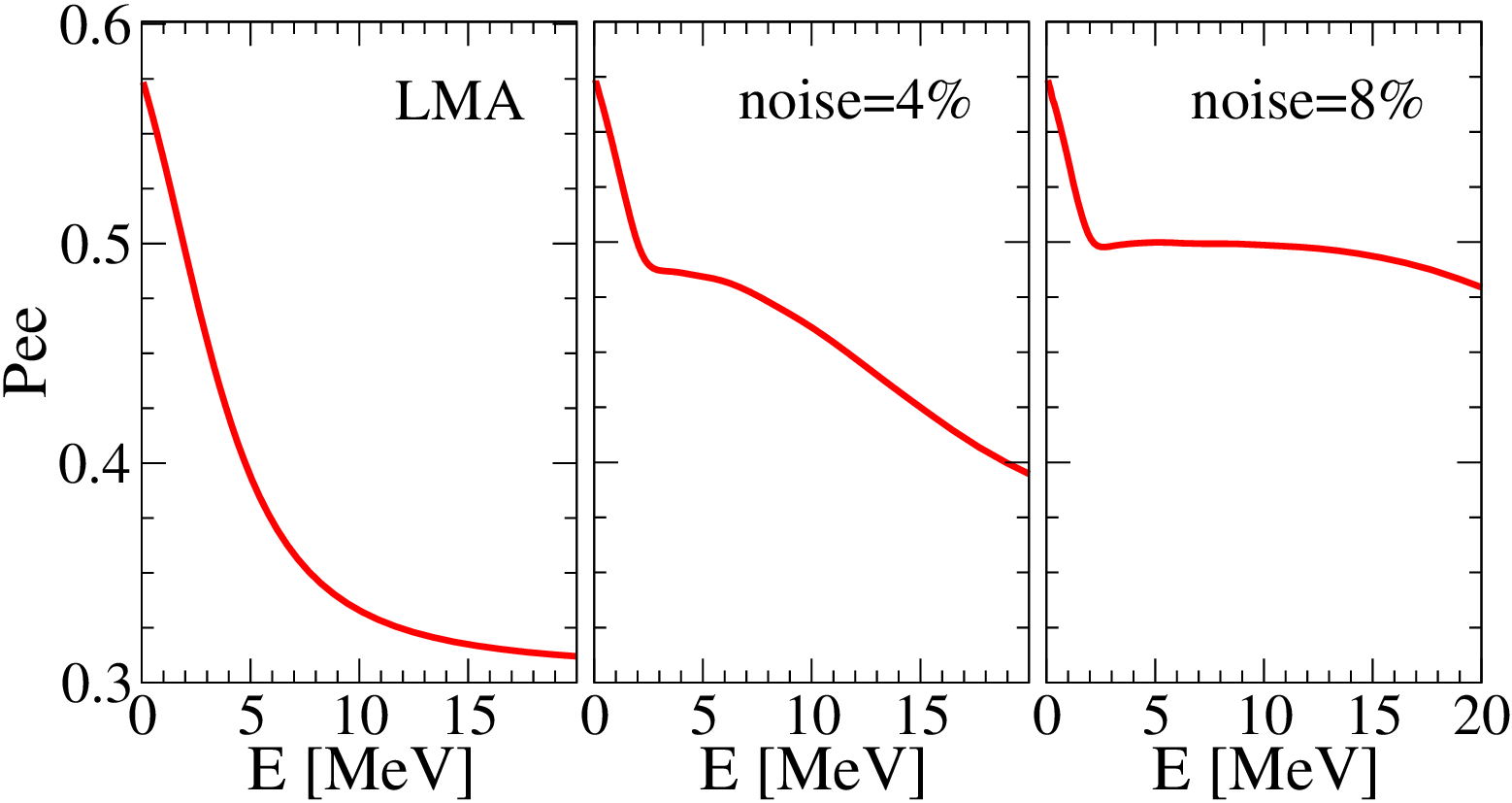}}
 \vspace{3mm}
 \caption{Effect of random electron density perturbations on
 electron-neutrino survival probability for LMA neutrino
 oscillations. The fluctuation's amplitude at the position of
 neutrino resonance, $\xi$, is zero
 in the left panel and is $\xi=4 \%$ and $\xi=8 \%$ in the
 middle and right panels, respectively. All panels use a
 fluctuation correlation length $L_0 = 100$ km.
 }
  \vglue-.5cm
    \label{fig:noisyLMA}
  \end{center}
\end{figure}

As fig.~\pref{fig:noisyLMA} shows, such fluctuations act to
degrade the efficiency of neutrino oscillations. They can do so
because successive neutrinos `see' slightly different solar
properties, and so in particular do not experience an equally
adiabatic transition as they pass through the neutrino resonance
region. The net effect is to degrade the effectiveness of the
neutrino conversion because those neutrinos for which the
transition is less adiabatic are more likely to survive as
electron-type neutrinos. Since the criterion for the transition to
be adiabatic depends on how quickly the electron distribution
varies near resonance, fluctuations give observable effects for
neutrinos if they occur at resonance with sufficient amplitude,
and if their correlation length, $L_0$, is comparable to the local
neutrino oscillation length, $L_{\rm osc} \sim 100$ km.

\subsection{The Implications of KamLAND and SNO Salt}
Ref.~\cite{usonSNO} has performed fits which are obtained using a
global analysis of the solar data, including radiochemical
experiments (Chlorine, Gallex-GNO and SAGE) as well as the latest
SNO data in the form of 17 (day) + 17 (night) recoil energy bins
(which include CC, ES and NC contributions,
see~\cite{Maltoni:2002ni})~\cite{Ahmad:2002jz} and the
Super-Kamiokande spectra in the form of 44
bins~\cite{Fukuda:2002pe}.

 \begin{figure}[htb!]
  \begin{center}
  \def\epsfsize#1#2{0.35#1}
  \centerline{\epsfbox{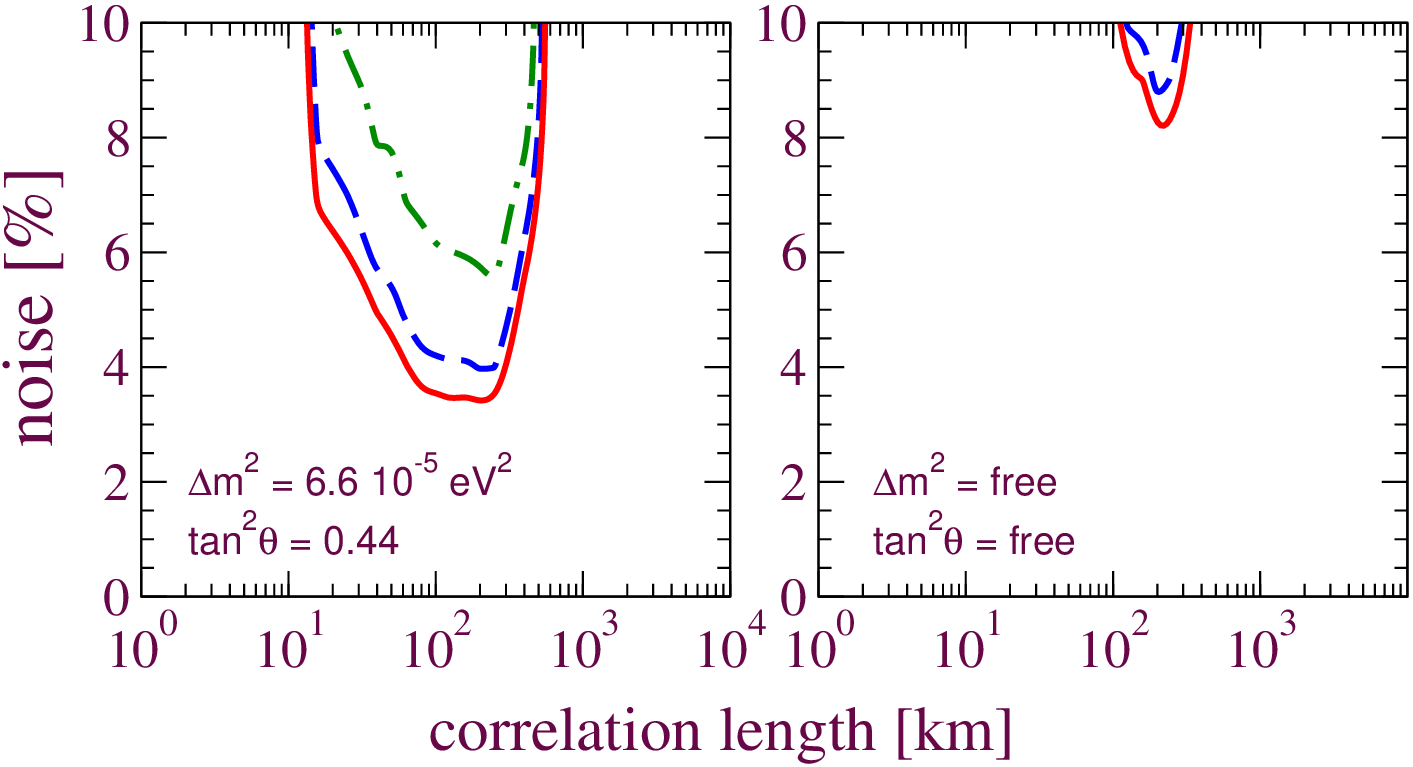}}
  \vspace{3mm}
    \caption{Exclusion region in the amplitude--correlation
    length ($\xi - L_0$) plane for solar fluctuations using only solar-neutrino data
    before the SNO salt measurements. In the left panel
    the neutrino oscillation parameters are assumed known
    while both oscillation and fluctuation parameters are
    jointly fit in the right panel. The lines indicate contours of
    90, 95 and 99\% CL.}
    \label{fig:chi2fit}
  \end{center}
\end{figure}
\vspace{-1cm}
 \begin{figure}[htb!]
  \begin{center}
  \def\epsfsize#1#2{0.5#1}
  \centerline{\epsfbox{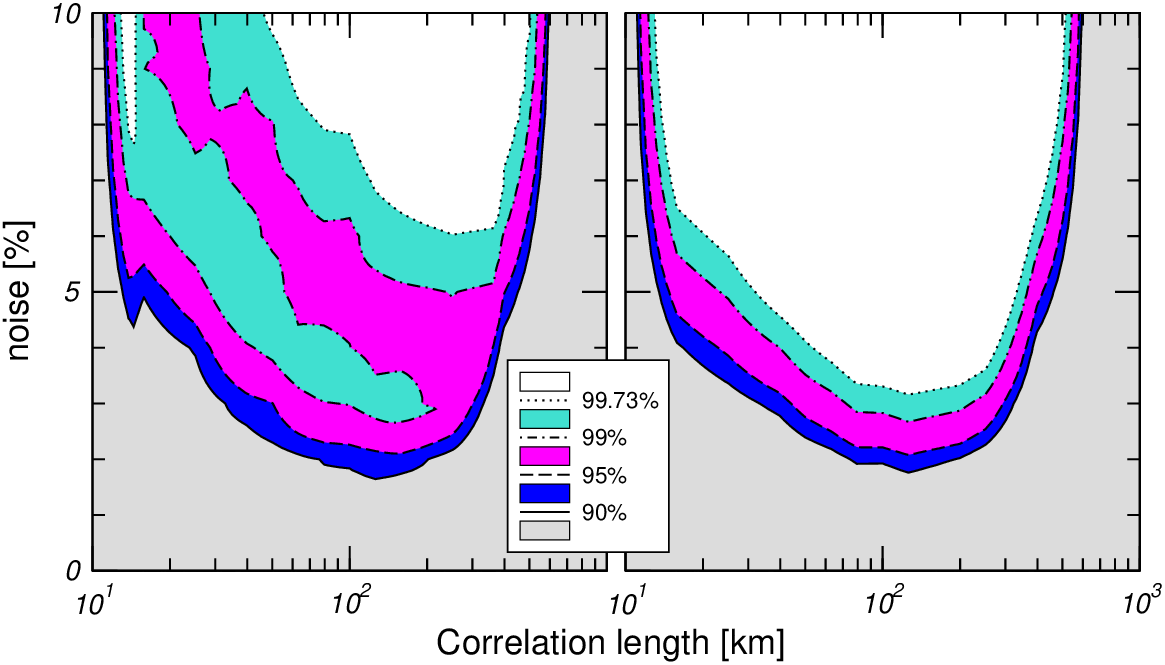}}
 \vspace{3mm}
 \caption{Sensitivity of solar neutrino data to the solar
    fluctuations including the SNO salt measurements. In this
    case the right panel
    assumes the neutrino oscillation parameters are known
    while the left panel shows the result when
    both oscillation parameters and
    fluctuations are jointly fit.}
  \vglue-.5cm
    \label{fig:noisyLMAsalt}
  \end{center}
\end{figure}

The sensitivity of the solar neutrino data to fluctuations in the
solar medium is summarized by figures \pref{fig:chi2fit} and
\pref{fig:noisyLMAsalt}. Fig.~\pref{fig:chi2fit} is taken from
ref.~\cite{us}, and summarizes the sensitivity before the SNO salt
measurements. Fig.~\pref{fig:noisyLMAsalt} gives the same results
after SNO salt. Comparing these figures shows the improvement in
constraints due to the SNO salt data, and comparing the panels in
each figure shows the the importance of a precise determination of
the neutrino oscillation parameters for obtaining a constraint on
the magnitude of fluctuations.

 \begin{figure}[htb!]
  \begin{center}
  \def\epsfsize#1#2{0.45#1}
  \centerline{\epsfbox{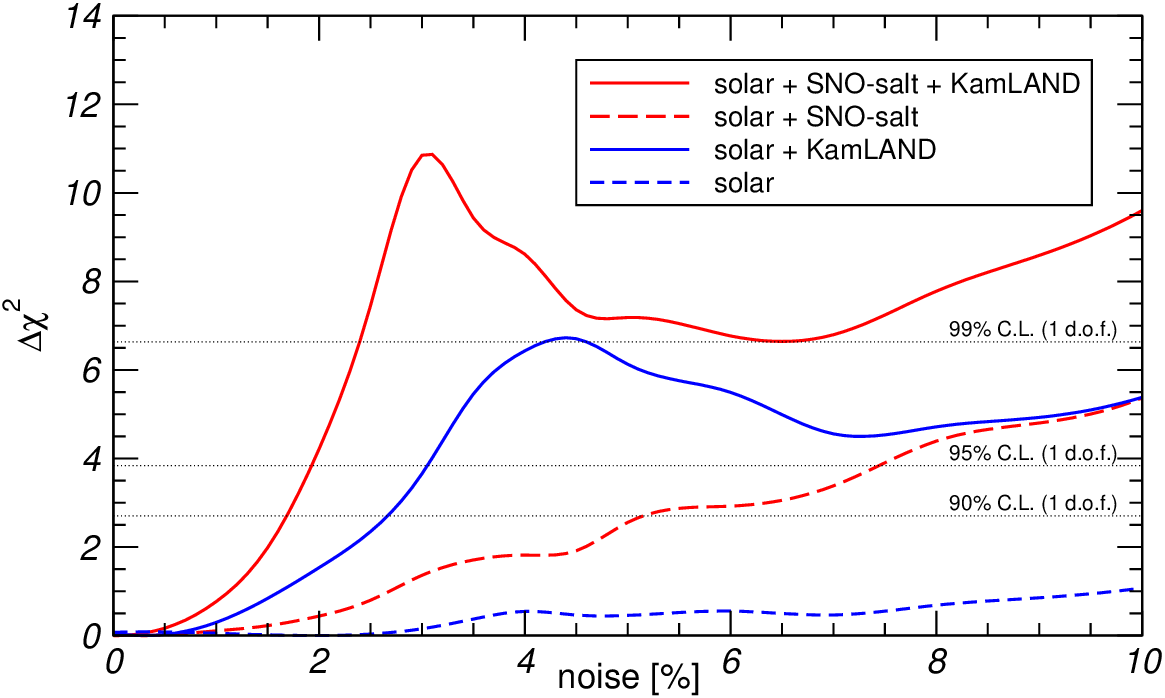}}
 \vspace{3mm}
 \caption{The chi-square of the fit as a function of fluctuation amplitude.}
  \vglue-.5cm
    \label{fig:noisyLMAchisq}
  \end{center}
\end{figure}
The importance of both the KamLAND and SNO salt measurements in
these results is most easily seen from
fig.~\pref{fig:noisyLMAchisq}, which compares the dependence of
the fit's $\chi^2$ on the amplitude of the fluctuations for
various data sets. This figure makes clear how the KamLAND results
are largely responsible for localizing the best fit near zero
fluctuation amplitude. This is as should be expected, since the
evidence for the absence of fluctuations follows from the
comparison of solar-neutrino observations with terrestrial
measurements of neutrino oscillation properties.

 \begin{figure}[htb!]
  \begin{center}
  \def\epsfsize#1#2{0.45#1}
  \centerline{\epsfbox{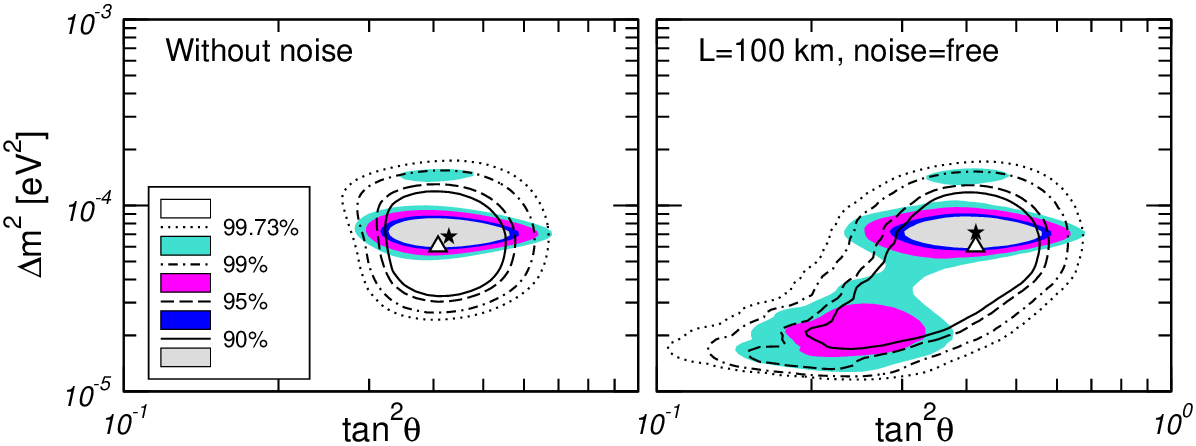}}
 \vspace{3mm}
 \caption{The solar-neutrino oscillation parameters obtained by the global fit.
 The results of the left panel are obtained assuming no noise, while those on the
 right fix the amplitude of the noise from the fit. The lines indicate
 confidence-level contours without using KamLAND data, while the coloured
 regions give the same information including KamLAND.}
  \vglue-.5cm
    \label{fig:LMAwwonoise}
  \end{center}
\end{figure}
Fig.~\pref{fig:LMAwwonoise} shows how the existence of solar
fluctuations influences the determination of the neutrino
oscillation parameters. The two panels of the figure contrast the
precision of the fit with and without solar fluctuations. The left
panel gives results subject to the usual prior assumption of no
solar fluctuations, while the right panel leaves the amplitude of
such fluctuations to be obtained from the fit. (When fluctuations
are included, they are assumed to have the optimal correlation
length $L_0 = 100$ km.) The lines indicate contours of fixed
confidence level when the KamLAND data is not included, while the
coloured regions give the same information when KamLAND is
included.

\begin{figure}[htb!]
  \begin{center}
  \def\epsfsize#1#2{0.45#1}
  \centerline{\epsfbox{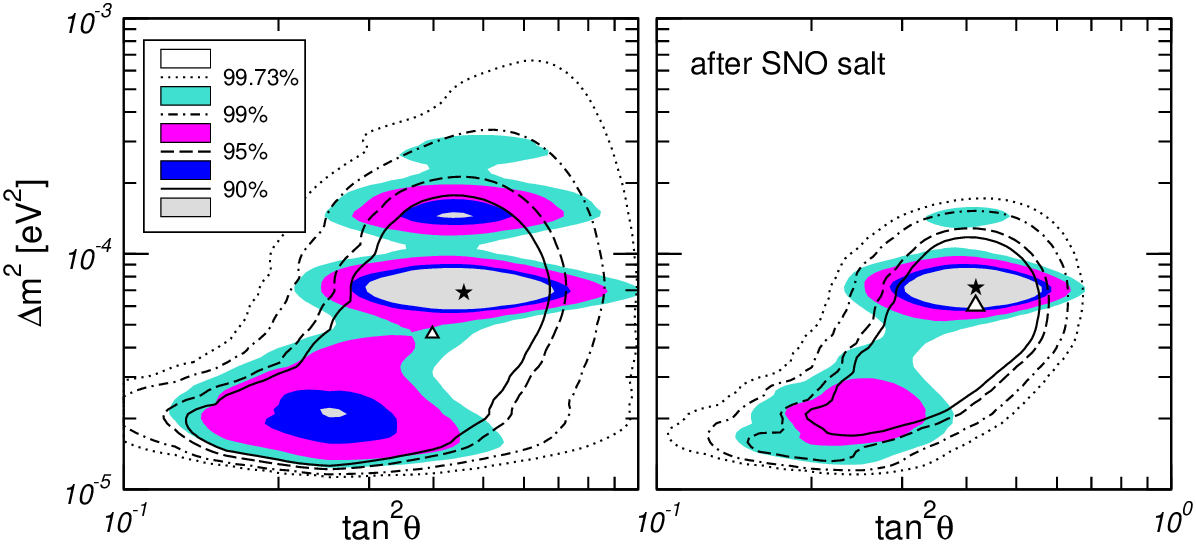}}
 \vspace{3mm}
 \caption{The same fit (including fluctuations) as above but
 performed with and without the SNO salt results.}
  \vglue-.5cm
    \label{fig:noisyLMAbeforesalt}
  \end{center}
\end{figure}

The main conclusion which follows from this figure is that the
precision with which the neutrino oscillations are known is now
largely independent of whether a prior assumption is made about
the existence of solar fluctuations. With the release of the SNO
salt results the comparison of solar-neutrino with KamLAND data
suffices to robustly determine the oscillation parameters
independent of the assumed amplitude of solar fluctuations. The
SNO salt data are crucial for reaching this conclusion, as is
clear from fig.~\pref{fig:noisyLMAbeforesalt}, which compares the
right panel of fig.~\pref{fig:LMAwwonoise} with the same fit
performed without using the SNO salt results.

We see that the SNO salt data, when combined with KamLAND results,
for the first time places the determination of
neutrino-oscillation parameters beyond the reach of sensitivity to
prior assumptions concerning the existence of fluctuations in the
solar radiative zone. Besides making more robust the determination
of neutrino-oscillation parameters, this allows a much crisper
determination of the kinds of solar fluctuations which can be
entertained deep within the solar radiative zone. As we have seen,
the resulting constraints apply to fluctuations whose spatial
scales are of order 100 km, and so are complementary to those
obtained from helioseismology, which are insensitive to
fluctuations on such short scales.

\section{Solar Magneto-Gravity Waves}

Given the experimental sensitivity to solar fluctuations
summarized above, the question remains as to whether a reasonable
source of fluctuations in the solar medium might exist having
sufficient amplitude and the correlation length required to
observably affect solar neutrinos. Ordinary helioseismic waves are
an obvious possibility, since they are known to exist deep within
the sun. Physically, they might have an effect because the waves
cause successive neutrinos to see different electronic density
profiles and so have the effect of making the neutrino `jump
probability' at the neutrino resonance into a random variable.

The influence of helioseismic waves on neutrino oscillations was
investigated in ref.~\cite{Bamert:1997jj}, where it was found they
are very unlikely to produce an observable effect. They do not for
one of two reasons. For those waves which definitely have been
observed ($p$-waves) the observed wave amplitude is much too small
deep inside the solar radiative zone to produce observable
effects. However, there are other helioseismic modes ($g$-modes)
which have not yet been observed but which must exist within the
solar radiative zone. It turns out that even if these modes are
assumed to have an amplitude as large as a few percent, their
wavelengths are too long to produce observable deviations from the
predictions of neutrino oscillations in the absence of solar
fluctuations.\footnote{A small number of potentially over-stable
modes could evade both of these arguments, but only if they were
to carry an inordinate amount of energy \cite{Bamert:1997jj}.}

The remainder of this review summarizes the results of
ref.~\cite{heliomag} which has suggested a possible mechanism for
obtaining observable fluctuations in the relevant part of the sun.
In this picture fluctuations having the appropriate distance scale
may arise if magnetic fields as large as of order 10 kG should
exist deep in the solar core. Magnetic fields of this size would
be consistent with current observational bounds
\cite{Couvidat:2002bs,Friedland}. It is not yet known whether such
modes could have sufficiently large amplitudes to allow observable
effects on neutrino oscillations, but the main lesson to be
learned from ref.~\cite{heliomag} is that the detailed shape of
helioseismic $g$-waves can be significantly changed by reasonable
radiative-zone magnetic fields, and so their influence on
neutrinos bears further study. The surprisingly strong sensitivity
to magnetic fields turns out to be driven by the occurrence of
level crossing between helioseismic $g$-modes and Alfv\'en waves.

Can 10 kG magnetic fields be present in the solar radiative zone?
Very little is directly known about magnetic field strengths there
-- see \cite{Cowling,Bahcall71} for early studies. A
generally-applicable bound is due to Chandrasekhar, and states
that the magnetic field energy must be less than the gravitational
binding energy: $B^2/8\pi \la GM^2_\odot/R_\odot^4$, or $B \la
10^8$ G. Stronger bounds are possible if one makes assumptions
about the nature and origins of the solar magnetic field. For
instance, if it is a relic of the primordial field of the
collapsing gas cloud from which the sun formed \cite{Parker}, then
it has been argued that central fields cannot exceed around 30 G
\cite{Boruta}. Similarly, the dynamo mechanism can only generate a
global field in the radiative zone of a newly-born Sun with
amplitude below 1 G \cite{Kitchatinov}. Even stronger limits, $B
\la 10^{-6}$ G apply~\cite{MestelWeiss} if the solar core is
rapidly rotating, as is sometimes proposed. On the other hand, it
has recently been argued \cite{Friedland} that fields up to 7~MG
could persist in the radiative zone for billions of years and are
consistent with current observational bounds. Other authors
\cite{Couvidat:2002bs} have recently entertained radiative-zone
fields as large as 30~MG.  Since the initial origin and current
nature of the central magnetic field is unclear, we consider as
admissible any magnetic field smaller than of order 10~MG.

\vskip 0.3cm \subsection{Magneto-gravity waves} \vskip 0.3cm
In this section we briefly review how magnetic fields influence
the equations of hydrostatic equilibrium on which helioseismic
analyses are based, and the approximations used to solve them.

The first relevant equations express conservation of mass:
\begin{equation}
\frac{d\rho}{dt} + \rho u = 0~ , \label{mass}
\end{equation}
where $d/dt = \partial/\partial t + \vector{v}\cdot \nabla$ is the
usual convective derivative, with $\vector{v}$ representing the
fluid velocity, and $\rho$ and $p$ are the fluid's mass density
and pressure. The variable $u = \div \vector{v}$ need not vanish
if the fluid is compressible.

The second equation of interest expresses energy conservation:
\begin{equation}
\frac{dp}{dt} - \gamma\frac{p}{\rho}\frac{d\rho}{dt} = - (\gamma -
1)Q~, \label{pressure}
\end{equation}
where $\gamma = c_p/c_V$ is given by the ratio of heat capacities,
and $Q$ is the sum of all energy density sources and losses, such
as heat conductivity, viscosity and and ohmic dissipation.

The magnetic fields enter more directly into the Euler equation
(conservation of momentum), which is of the form
\begin{equation}
\rho \frac{d\vector{v}}{dt} = - \nabla p  + \rho \vector{g} +
\frac{1}{4\pi }\Bigl [\rot \vector{B}\times \vector{B}\Bigr ]~,
\label{Euler}
\end{equation}
with the local force of gravity given by $\rho \vector{g}$. The
local acceleration due to gravity is related to the Newtonian
potential by $\vector{g} = \nabla \phi$, with $\phi$ given by the
Poisson equation $\nabla^2 \phi = - 4\pi G\rho$. As usual, $G$
here denotes Newton's constant. The last term of equation
\pref{Euler} expresses the contribution of the Lorentz force to
the local momentum budget.

Finally, the system is completed by Faraday's equation,
\begin{equation}
\frac{\partial \vector{B}}{\partial t} = \rot [\vector{v}\times
\vector{B}] + \frac{c^2}{4\pi \sigma_{cond}} \nabla^2 \vector{B}~,
\end{equation}
that governs the time evolution of the magnetic field. Here
$\sigma_{cond}$ represents the fluid's conductivity. In what
follows we take the plasma to be an ideal conductor, meaning we
take $\sigma_{cond}$ to be large enough to neglect the last term
in this last equation.

In order to find solutions to these equations a number of
assumptions are required. We list them all here for ease of
reference. In deriving the differential equations to be solved, we
adopt the following approximations.
\begin{enumerate}
\item
In principle, electromagnetic processes enter into
Eq.~\pref{pressure} through their contributions to $Q$, but for
ideal MHD we neglect both the heat conductivity and viscosity
contributions to energy losses, as well as the ohmic dissipation,
$Q = j^2/\sigma_{cond}$, where $\sigma_{cond}$ is the electrical
conductivity.
\item
We linearize the equations about a static background
configuration, {\it i.e.} a background configuration which is time
independent and for which the background fluid velocity vanishes,
$\vector{v}_0 = 0$.
\item
We assume the thermodynamic quantity, $\gamma = c_p/c_V$, has
vanishing derivative for a Lagrangian fluid element, $d\gamma/dt =
\partial \gamma/\partial t + {\bf v} \cdot \nabla \gamma =  0$.
This would be true, in particular, for a polytropic gas.
\item
We adopt the Cowling approximation, which amounts to the neglect
of perturbations of the gravitational potential, ({i.e.:} $\phi' =
0$).
\item
We assume the fluctuations to be adiabatic, with the contributions
of fluctuations to the heat source vanishing: $Q'=0$ (this is
satisfied with good accuracy by high frequency oscillations).
\item
We assume no background electric or displacement currents, so the
background magnetic field satisfies $\rot \vector{B}_0 = 0$.
\end{enumerate}

Finally, we are able to solve the resulting equations analytically
if we make two further assumptions.
\begin{enumerate}
\item[7.]
We assume a rectangular geometry, with background quantities
varying along the $z$ direction (which implies the local
gravitational acceleration, ${\bf g}$, is directed along the $z$
axis). We also take a constant, uniform background magnetic field,
${\bf B}_0$, pointing along the $x$ axis.
\item[8.]
The background mass-density profile is assumed to be exponential,
$\rho_0 = \rho_c \, \exp[- z/H]$, for constant $\rho_c$ and $H$.
The conditions of hydrostatic equilibrium for the background then
determine the profiles of thermodynamic quantities, and in
particular imply $\gamma$ is a constant.
\end{enumerate}

This last approximation applies to very good approximation for
real mass-density profiles obtained by standard Solar models,
provided we identify the $z$ direction with the radial direction,
and focus our attention to deep within the radiative zone. The
constancy of $\gamma$ in this region is also expected since the
highly-ionized plasma satisfies an ideal gas equation of state to
good approximation. The rectangular geometry provides a reasonable
approximation so long as we do not examine too close to the solar
centre. What is important about our choice for ${\bf B}_0$ is that
it is slowly varying in the region of interest, and it is
perpendicular to both ${\bf g}$ and all background gradients,
$\nabla \rho_0$, $\nabla p_0$, {\it etc}.

\subsection{The eigenvalue problem}
Using the above assumptions we can express all oscillating
quantities in terms of any one component, which we take to be the
profile $b_z(z)$. This quantity turns out to satisfy a linear
ordinary second-order differential equation whose solution
determines all of the other variables through simple expressions
\cite{heliomag}. The relevant ordinary differential equation is
\begin{equation}
\left(1 - \frac{k_x^2v_A^2}{\omega^2}\right)\frac{d^2b_z(z)}{dz^2}
- \frac{N^2}{g}\frac{db_z(z)}{dz} + k_{\perp}^2\left(
\frac{k_x^2v_A^2}{\omega^2} - 1 + \frac{N^2}{\omega^2}
\right)b_z(z) = 0~, \label{masterbz}
\end{equation}
where $N$ denotes the Brunt-V\"ais\"al\"a frequency, as defined by
\begin{equation}
N^2(z)= g(z)\left(\frac{1}{\gamma \, p_0} \, \frac{dp_0}{dz} -
\frac{1}{\rho_0} \, \frac{d\rho_0}{dz} \right) ,
\end{equation}
and $v_A = B_0/\left(4 \pi \rho_0\right)^{1/2}$ defines the
Alfv\'en speed. $k_x$ and $k_y$ here denote the wave-number in the
transverse directions. Equation~\pref{masterbz} contains all of
the information required about the system's normal modes. Notice
that its derivation allows the quantities $g$, $\gamma$, $\rho_0$
and $p_0$ to depend generally on $z$, subject only to the
conditions of hydrostatic equilibrium and vanishing Lagrangian
derivative, $d\gamma/dt = 0$.

The boundary conditions to be imposed are as follows. At the solar
center ($z=0$) we use the boundary condition which would have
arisen as a smoothness condition in cylindrical coordinates:
\begin{equation} b_z(0)=
k_x v_z(0)/\omega = 0 .
\end{equation}
The other boundary condition is imposed at the top of the
radiative zone, $z_* \approx 0.7 R_\odot$. In this region the two
solutions to the differential equation behave as $e^{\pm k_\perp
\, z}$ (where $k_\perp^2 = k_x^2 + k_y^2$) and so either fall or
grow exponentially as functions of $z$. We take the absence of the
growing behaviour to be our boundary condition for $z=z_*$.

\begin{figure}
\includegraphics[width=\columnwidth]{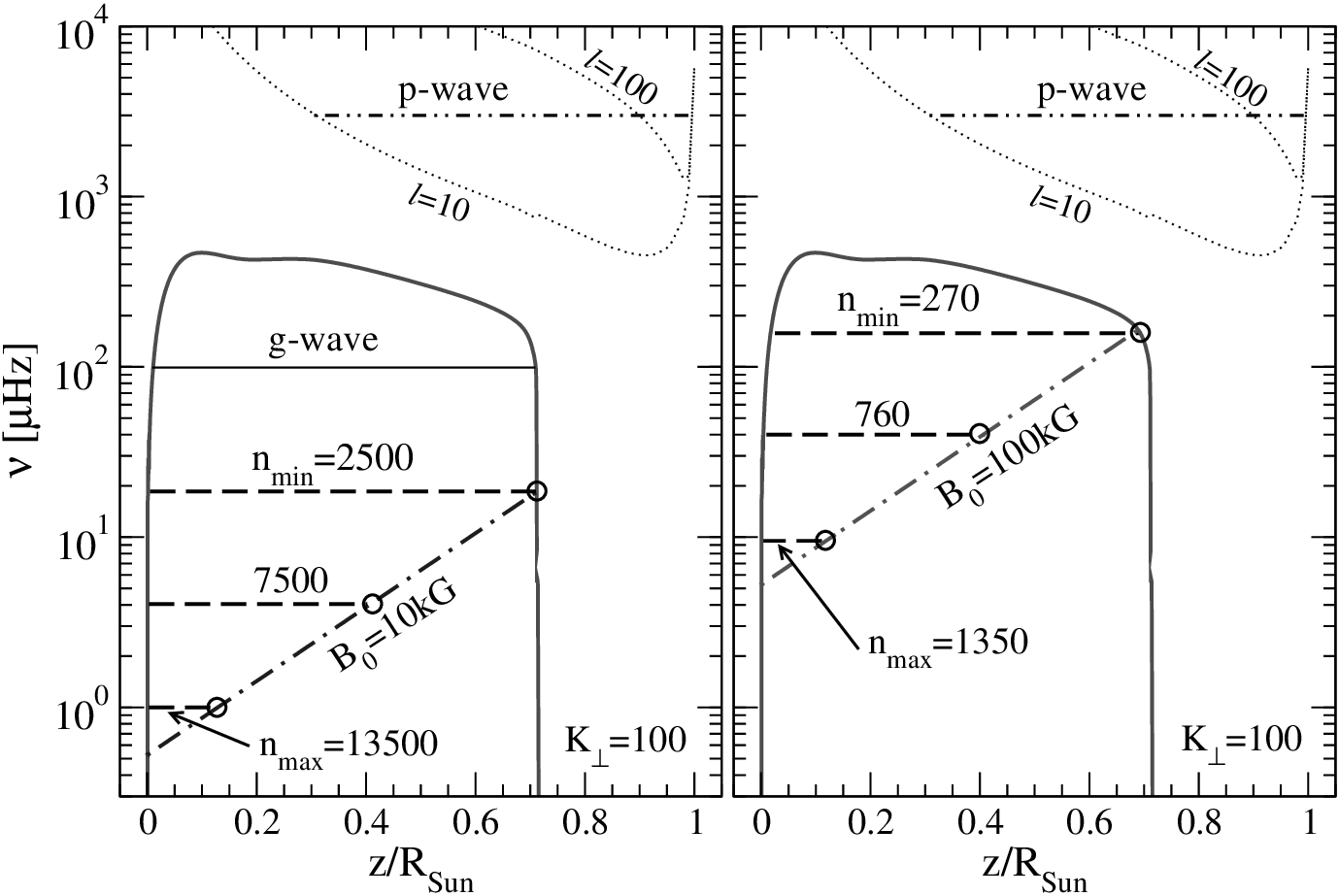}
\caption{Relevant frequencies  plotted against radial position
within
  the sun. The solid curve gives the Brunt-V\"ais\"al\"a frequency,
  $N(z)$, while the dot-dashed lines are the Alfv\'en frequencies.
  Notice that $N(z)$ goes to zero at the solar center and the top of
  the radiative zone. The horizontal dashed lines represent the
  frequencies of magneto-gravity waves
  for several choices for the mode number, $n$, for fixed values
  $k_x
  = k_\perp = 100/H$. As discussed in the text, resonances occur where
  the mode frequencies intersect the Alfv\'en frequency, indicated by
  circles in the figure. The two panels correspond to two choices for
  the magnetic field: 10~kG (left) and 100~kG (right). Minimum and
  maximum mode numbers are indicated, with $n_{\rm min}$ defined by
  the condition that its resonance occurs at the top of the radiative
  zone, $z_r(n_{\rm min})=0.7R_\odot$, and $n_{\rm max}$ having
  resonance at $z_r(n_{\rm max})=0.12R_\odot$. The dotted lines
  denote acoustic (Lamb) frequencies for $l=10$ and $100$ (see
  \cite{Christensen-Dalsgaard:2002ur}) while the double-dot-dashed
  horizontal line represents the trapping region for a p-wave of
  frequency 3000~$\mu$Hz and $l=10$. The horizontal solid line
  represents the trapping region for a g-wave of frequency
  100~$\mu$Hz.
\label{bruntfig1}}
\end{figure}

It is instructive to examine the qualitative properties of the
eigenvalue problem we have obtained. In the limit $B_0 \to 0$ (and
so $v_A \to 0$), Eq.~(\ref{masterbz}) reduces to the standard
evolution equation for `pure' helioseismic $g$-modes, which is
usually expressed in terms of the variable $v_z$ (which is related
to $b_z$ through the relation $v_z=\omega b_z/k_x$). Since more
detailed analyses -- see Fig.~\ref{bruntfig1} -- show that $N$
rises from zero at the solar center, remains approximately
constant $N \approx N_0$, through the radiative zone, and then
falls to zero again at the bottom of the convective zone, these
$g$-modes can be thought of as the eigen-modes of oscillations
inside the cavity in between the two regions where $N$ goes
through zero. For a given wave frequency, $\omega < N_0$, the
turning points of this cavity are given by the condition $\omega
\approx N$. For smaller $\omega$ the lower turning point gets
closer to the solar center, $z\to 0$, and the upper one gets
slightly closer to the bottom of the convective zone (CZ).

Conversely, if gravity is turned off ($N \to 0$), then
Eq.~(\ref{masterbz}) describes Alfv\'en waves, which oscillate
with frequency $\omega = k_x v_A$ and propagate along the magnetic
field lines. Notice that since $v_A \propto \rho^{-1/2}$ this
frequency grows with $z$, since the density of the medium falls.

Keeping both magnetic and gravitational fields introduces
qualitatively new behavior, as may be seen mathematically because
equation (\ref{masterbz}) acquires a new singular point which
occurs when the coefficient of the second-derivative term
vanishes. Since this singularity appears at the Alfv\'en
frequency,
\begin{equation}
  \label{eq:reson1}
\omega =k_xv_A
\end{equation}
it can be interpreted as being due to a resonance between the
$g$-modes and Alfv\'en waves. This resonance turns out to occur at
a particular radius because the Alfv\'en frequency varies with
radius while the $g$-mode frequency is independent of radius (see
Fig.~\ref{bruntfig1}). The resonance occurs where the growing
Alfv\'en mode frequency crosses the frequency of one of the
$g$-modes, and the resulting waveforms would be expected to vary
strongly at these points. The resonance can occur inside the
radiative zone if the Alfv\'en frequency climbs high enough to
cross a $g$-mode frequency before reaching the top of the
radiative zone. Since $v_A \propto B_0$, whether this occurs or
not depends on the field value, $B_0$.

The qualitative behaviour of the solutions can be seen from
equation (\ref{masterbz}), and are oscillatory (or exponentially
growing or damped) if the coefficient of the last term of this
equation is positive (or negative). In the absence of magnetic
fields, this leads to the onset of damping when $\omega > N(z)$.
Since $N(z)$ vanishes at the top of the radiative zone, all modes
necessarily become unstable there, and the corresponding
instability gives rise to the convection which defines the
convective zone. This is the usual picture of helioseismic $g$
modes. The presence of magnetic fields makes the coefficient of
$b_z$ in eq.~(\ref{masterbz}) go negative (for some modes) at
smaller values of $z$, indicating the onset of an instability
inside of the radiative zone.

The magnetic field leads to several new effects. First, the
shortening of the cavity due to the presence of magnetic field
causes the eigen-frequencies to depend on the magnetic field
value. Second, there can be energy transfer between $g$-modes and
Alfv\'en waves, within the narrow singular resonance layer,
leading to the corresponding eigen-frequencies acquiring imaginary
parts. Third, the nearer the upper boundary of MHD cavity is to
the solar center, the stronger the $g$-modes are confined to the
solar core.

More details of these waveforms can be obtained from a WKB-type
analysis of the master Eq.~(\ref{masterbz}). Near the solar center
where $N^2\to 0$, if $(k_xv_A)^2/\omega^2\ll 1$ one obtains the
exponential solution $b_z\to e^{\pm k_{\perp}z}$. (Which
combination of these solutions appears is fixed by boundary
conditions, {e.g.} if $b_z(0) = 0$ at the solar center we have
$b_z \propto \sinh(k_\perp z)$.) One finds similar behaviour for
the solution above the singular resonant layer, $z>z_r$, up to the
top of the radiative zone. This exponential growth happens because
of the exponential decrease of density (and so exponential
increase in $v_A$) with $z$. The requirement for complex
frequencies arises from the demand that the solution remain
regular in the narrow Alfv\'en resonance layer.

\begin{figure}
\includegraphics[width=\columnwidth]{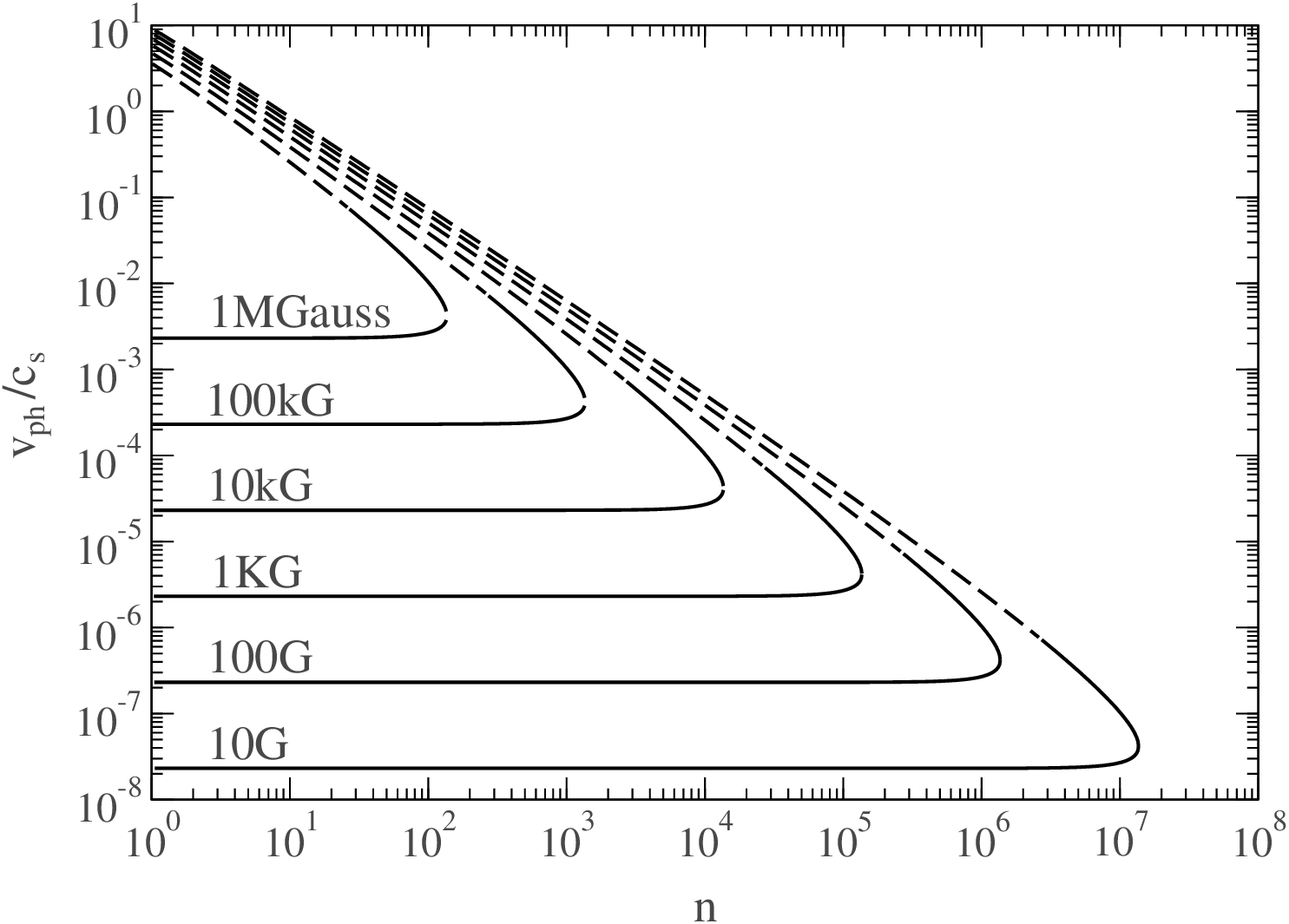}
\caption{Plot of the $v_{ph}/c_s = \omega_1/k_x c_s$ against mode
number, $n$, where $c_s$ is the adiabatic sound speed and
$\omega_1$ is the real part of the eigenfrequencies, $\omega =
\omega_1 (1 + i d)$. Different curves correspond to different
background magnetic field strengths, ${B}_0$. Solid lines
represent resonances which are inside the radiative zone, whilst
dashed lines correspond to unphysical modes whose resonances lie
outside of the radiative zone.\label{spectrumfig1}}
\end{figure}

\begin{figure}
\includegraphics[width=\columnwidth]{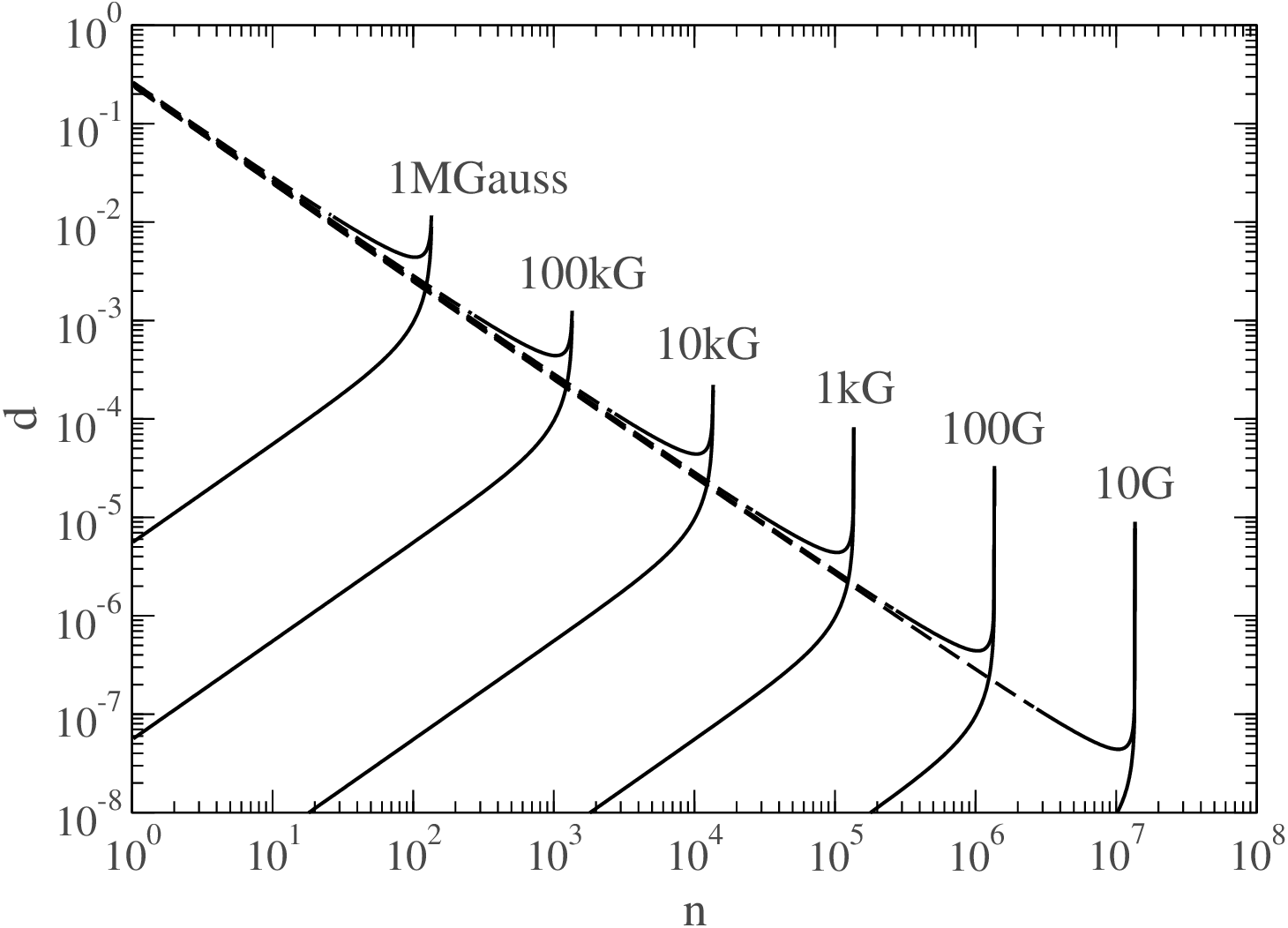}
\caption{Plot of the $d$ against mode number, $n$, where $d$ gives
the imaginary part of the eigenfrequencies, $\omega = \omega_1 (1
+ i d)$. Different curves correspond to different background
magnetic field strengths, ${B}_0$. Solid lines represent
resonances which are inside the radiative zone, whilst dashed
lines correspond to unphysical modes whose resonances lie outside
of the radiative zone.\label{spectrumfig2}}
\end{figure}

Figures~\ref{spectrumfig1} and~\ref{spectrumfig2} present our
numerical solution of the eigenvalue spectrum.
Figure~\ref{spectrumfig1} plots $v_{ph}(n)/c_s = \omega_1/(c_s
k_x)$ {\it vs} mode number $n$ for various magnetic fields, $B_0$.
Here $\omega_1$ is the real part of the eigenfrequency $\omega$,
which in general is complex. (More about this later.) Figure
\ref{spectrumfig2} similarly plots the imaginary part of omega, Im
$\omega = \omega_1 \, d$ against mode number for the same magnetic
fields. In both plots the parameter $\alpha=k_x/k_{\perp}$ is
taken to be unity, where as before $k_x$ and $k_\perp^2 = k_x^2 +
k_y^2$ are the mode's wave-numbers in the transverse directions.
In both figures a dashed line is used if the resonance of interest
occurs above the top of the radiative zone (and so outside the
domain of many of our approximations).

Both of the figures~\ref{spectrumfig1} and~\ref{spectrumfig2} show
two branches of solutions up to a maximum mode number, as
expected. Their dependence on $n$ can also be understood
analytically from following approximate expressions
\cite{heliomag}
\begin{eqnarray}
\label{superlow} &&\frac{\omega _{1}}{N}=\frac{2 k_\perp H}{\pi n}
\ln\left(\frac{4NH}{\alpha v_{Ac}}\frac{1}{\pi n}\right),\nonumber\\
&&d=\frac{\ln \left|\tan (\pi/\gamma )\right|}{2\pi n}~.
\end{eqnarray}
These modes correspond to the branches of the
figures~\ref{spectrumfig1} and~\ref{spectrumfig2} for which both
$v_{ph}/c_s$ and $d$ fall with $n$. Analytic expressions are also
possible for the other branch \cite{heliomag}, and the spectrum in
this case is
\begin{eqnarray}
    &&\omega _{1}=\frac{\alpha K_{\bot } v_{Ac}}{H},\nonumber\\
    &&d=-\left( \frac{\alpha v_{Ac}}{4NH}\right) ^{2}  \pi n \ln \left| \tan
    \frac{\pi }{\gamma }\right|~.
\end{eqnarray}
Note that for this branch $v_{ph}(n)/c_s$ is independent of the
mode number, $n$, and $d$ grows with $n$, as is also seen in the
figure.

In addition to requiring the resonance to occur in the radiative
zone (the solid line in the figures), the validity of our
approximations also demand the frequency not to be smaller than
$10^{-5}$~s$^{-1}$ due to our neglect of the 27-day period solar
rotation. Inspection of Fig.~\ref{spectrumfig1} shows that these
two conditions are consistent with one another for a reasonably
large range of modes only for magnetic fields larger than a kG or
so.

The resonance alluded to above appears causes two distinctive
features to appear in these solutions. First, the eigenfrequencies
are complex, implying both damped and exponentially-growing modes.
Second, the eigenmodes are not smooth as functions of $z$ about
the singular resonant point, $z=z_r(n,k_x,k_y)$, whose position
depends on the quantum numbers of the mode in question.

The necessity for complex frequencies imply the mode functions
grow exponentially in amplitude with time. This signals an
instability in the physics which pumps energy into these
resonances, and this section aims to discuss the nature of this
instability, and the physical interpretation to be assigned to the
imaginary part, $d$. Exponentially-growing instabilities within an
approximate ({\it e.g.} linearized) analysis reflect the system's
propensity to leave the small-field regime, on which the validity
of the approximate analysis relies. The question becomes: where
does the instability lead, and what previously-neglected effects
stabilize the runaway behaviour?

In the present instance the normal modes are strongly peaked near
the resonant radii, and the energy flow near these radii is
directed along the resonance plane, much as would be true for a
pure Alfv\'en wave. Since helioseismic waves are likely generated
by the turbulence at the bottom of the convective zone, it is
natural to imagine starting the system with a regular helioseismic
$g$-mode and asking how it evolves. In this case the resonance
allows the energy in this mode to be funnelled into the Alfv\'en
wave, and so to be channelled along the magnetic field lines away
from the solar equatorial plane. The imaginary part of the
frequency, $\Gamma = \hbox{Im} \, \omega =  \omega_1 \, d$,
describes the rate at which the Alfv\'en mode is excited in this
process.

Once excited, the mode amplitudes near the resonances grow until
the energy in them becomes dissipated by effects which are not
captured by the approximate discussion we present here. The rate
for this dissipation must grow as the mode amplitude grows, until
it equals the production rate, $\Gamma$, at which point a steady
state develops and the mode growth is stabilized. Because the
resonant mode grows until its damping rate equals its production
rate, it suffices to know the mode's growth rate in order to
determine the overall on-resonance amplitude. If the mode
stabilizes once it is large enough to require a nonlinear
analysis, then the final production and dissipation rates may be
very different from the linearized expressions derived above. If,
on the other hand, the modes saturate at comparatively small
amplitudes by dissipating energy into non-hydrodynamical modes
(such as by Landau damping), then it can happen that the
stabilized mode amplitude is not large enough to significantly
change the linearized prediction for its production rate.

\begin{figure}
\includegraphics[width=\columnwidth]{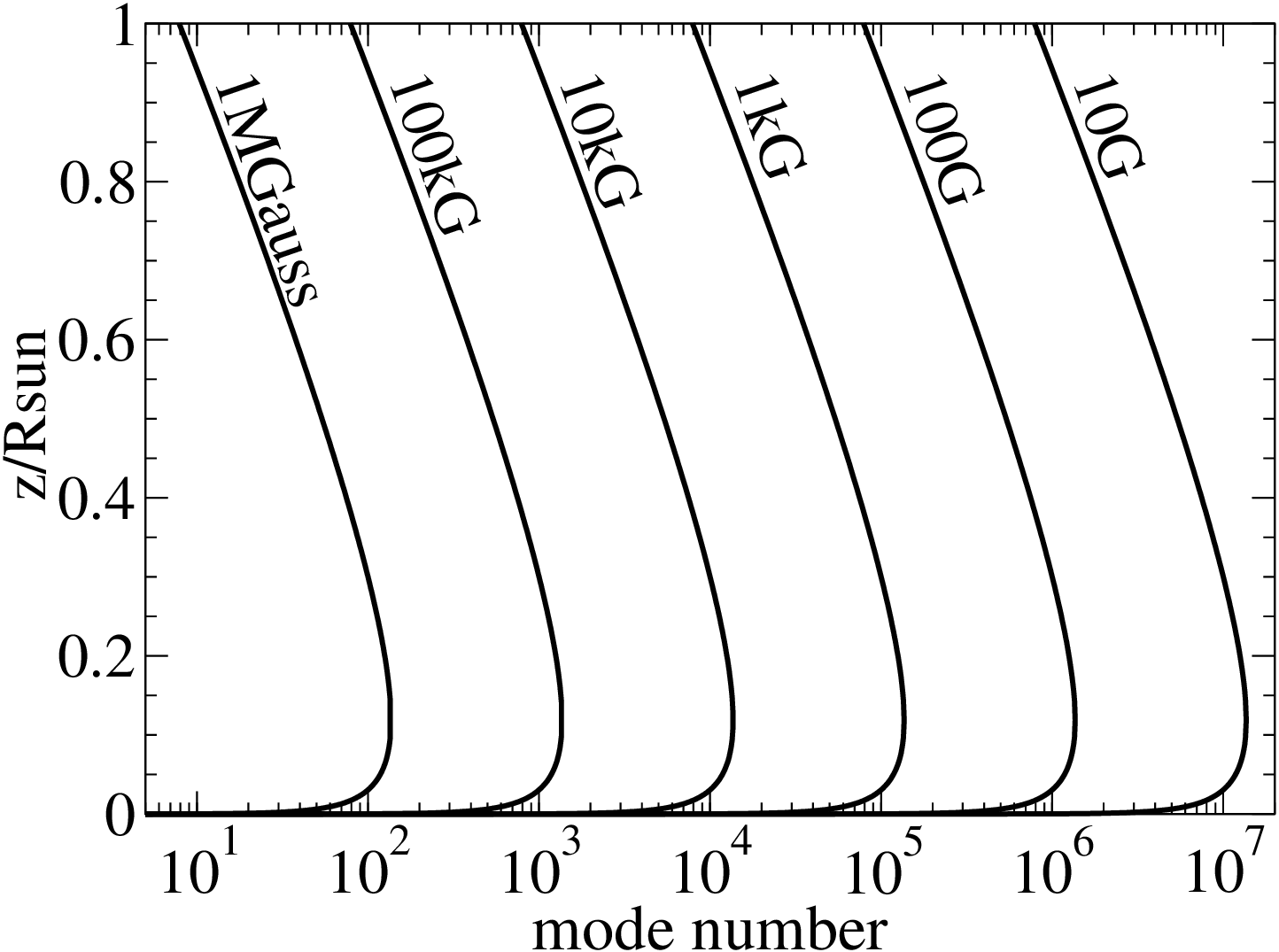}
\caption{A plot of the resonant position, $z_r(n)$, {\it vs} mode
number $n$ for different magnetic fields in the range $B_0
=$~10G--1MG.\label{resposfig}}
\end{figure}

\begin{figure}
\includegraphics[width=\columnwidth]{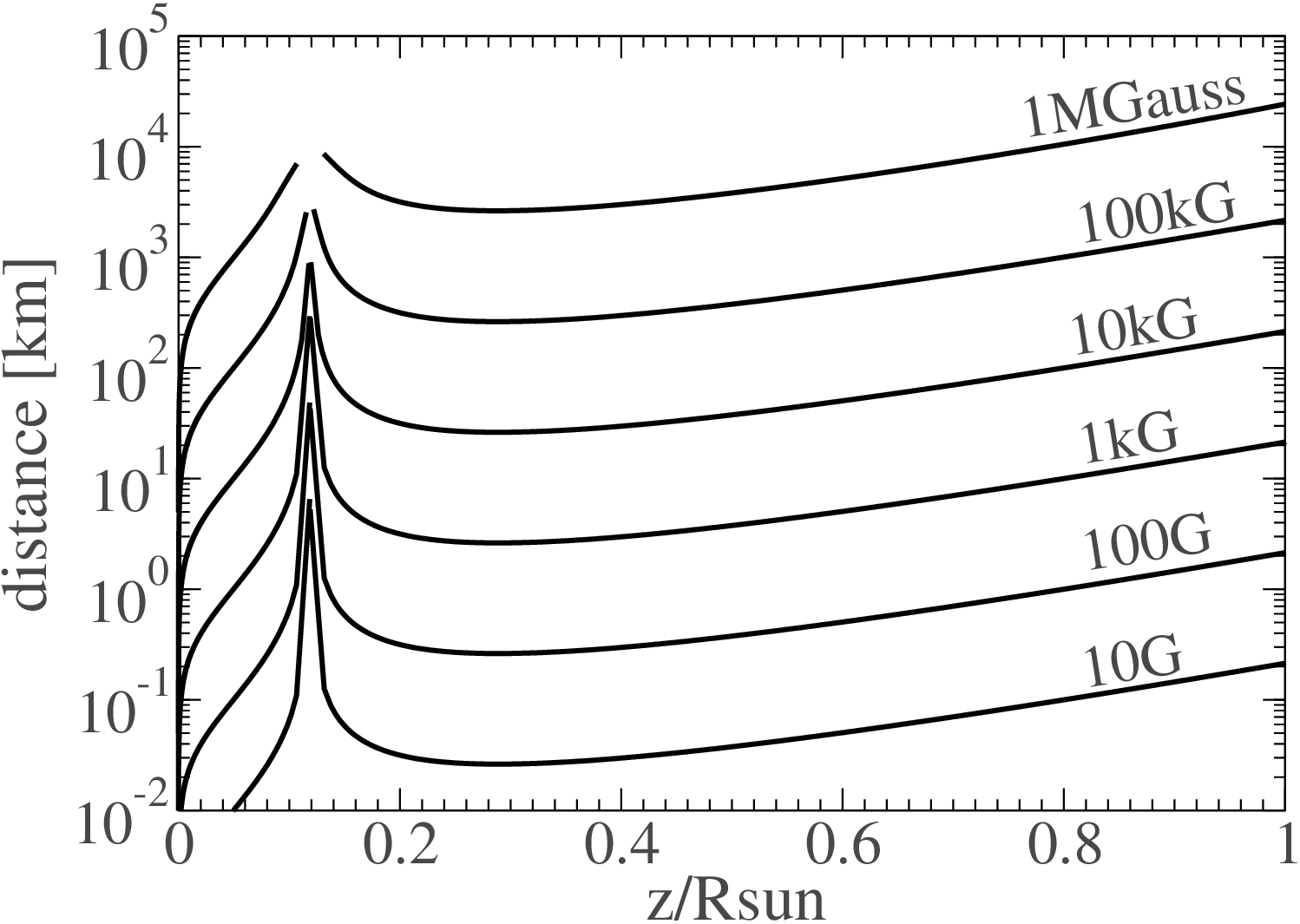}
\caption{The distance between neighbouring Alfv\'en resonant
layers {\it vs} the position of the layer within the solar
interior. \label{resspacefig}}
\end{figure}

Fig.~\ref{resposfig} plots the position, $z_r(n)$, predicted for
the case of longitudinal wave propagation ($k_y = 0$) and for
magnetic fields in the range $B_0=$10G--1MG. (The same result for
an obliquely-propagating wave with $k_y \ne 0$ is produced by a
larger value for $B_0$.) This plot is the analog of Figs.~5a,b
in~\cite{DzhalilovSemikoz}. Knowing the position of the resonances
also permits us to determine the distance between them. This
quantity is relevant to the propagation of particles like
neutrinos through the resonant waves. The spacing is:
\begin{equation}
    z_r(n +1) - z_r(n) = \left(\frac{k_x}{k_{\perp}}\right) \frac{\pi
    H \, v_{Ac}\gamma}{c_s\sqrt{\gamma -1}} \; e^{z_r(n)/2H} ~.
\label{distance}
\end{equation}
This dependence of this quantity on mode number, $n$, is shown in
Fig. \ref{resspacefig} for longitudinal wave propagation, $k_y=0$,
and for different magnetic field values. From this figure we see,
in the region $z_r\ga 0.3R_{\odot}$, that the distance between
layer positions grows with magnetic field and with distance from
the solar center. Both of these features were already seen in
preliminary WKB calculations of the resonances (see Figs. 7a, 7b
of ref.~\cite{DzhalilovSemikoz}), and our present numerics also
confirm that the spacing is approximately proportional to $B_0$
for $z \ga 0.3 R_\odot$, that was seen earlier.

An approximate expression for the differenece between adjacent
resonance positions is given by
\begin{equation}
\label{simple} z_r(n +1) - z_r(n) \approx \frac{2H}{\mid n\mid}~,
\end{equation}
where the inverse mode number $\mid n \mid^{-1}$ is proportional
to the background magnetic field $\sim B_0$.

Numerically, it is noteworthy that the spacing between resonances
can be hundreds of kilometers. This is significant because this is
close to the resonant oscillation length, $l_{osc}^{res}$, for
$E\sim \rm{MeV}$ neutrinos, if -- as now seems quite likely --
resonant LMA oscillations are responsible for explaining the solar
neutrino problem since
\begin{equation}
\label{nuosc} l_{osc}^{res}= \frac{250~\rm{km}~
(E/\rm{MeV})}{\Delta m^2_5\sin 2\theta}\,,
\end{equation}
where $\Delta m^2_5 = \Delta m^2/10^{-5}~ \hbox{eV}^2$. Repeatedly
perturbing neutrinos over distance scales comparable to
$l_{osc}^{res}$ has long been known to be a prerequisite for
disturbing the standard MSW picture of oscillations in the solar
medium. This raises the possibility -- recently explored in more
detail in ref.~\cite{us} -- that $g$-mode/Alfv\'en resonances can
alter neutrino propagation. If so, the observation of resonant
oscillations of solar neutrinos may provide some direct
information about the properties of the MG waves we discuss here.

\section{Summary}
Within the approximations given it appears that sufficiently large
magnetic fields can cause significant changes to the profiles of
helioseismic $g$-waves, while not appreciably perturbing
helioseismic $p$-waves. The comparatively large $g$-wave effects
arise because of a resonance which occurs between the $g$-modes
and magnetic Alfv\'en waves in the solar radiative zone. Although
the radiative-zone magnetic fields required to produce observable
effects are larger than are often considered -- more than a few kG
-- they are not directly ruled out by any observations.

Although the density profiles at their maxima could have
amplitudes as large as a few percent or more on resonance, we do
not believe the corresponding radiative-zone magnetic fields can
yet be ruled out by comparison with helioseismic data, since the
density excursions are sufficiently narrow (hundreds of
kilometers) as to evade the assumptions which underlie standard
helioseismic analyses.

For magnetic fields in the 10 kG range, the best hopes for
detection of the resonant waves may be through their influence on
neutrino propagation. This influence essentially arises because
the presence of strong density variations affects the solar
neutrino survival probability, with a corresponding change in the
resulting solar neutrino fluxes.  As described above, and shown in
ref.~\cite{us}, the measurement of neutrino properties at KamLAND
provides new information about fluctuations in the solar
environment on correlation length scales close to 100~km, to which
standard helioseismic constraints are largely insensitive. We have
already seen how the determination of neutrino oscillation
parameters from a combined fit of KamLAND and solar data depends
strongly on the magnitude of solar density fluctuations.

Since the resonances rely on the condition that $\vector{B} \perp
\nabla \rho$, there are several magnetic-field geometries to which
our analysis might apply, and it is instructive to consider two
illustrative examples to see what kinds of observable effects
might be possible. Suppose first that, in spherical coordinates
$(r,\theta,\phi)$, we imagine $B_r \approx 0$ but $B_\theta \ne
0$. Then the field is always perpendicular to a radial density
gradient and the resonance we find might be expected to arise in
all directions as one comes away from the solar center. In this
case the solar $g$-modes would tend to be trapped behind the
resonance, and so are kept away from the solar surface even more
strongly than is normally expected. This would make the prospects
for their eventual detection very poor.

Alternatively, if the magnetic field has more of a dipole form it
might be imagined that the condition $\vector{B} \perp \nabla
\rho$ only holds near the solar equatorial plane, and not near the
solar poles. In this case a more detailed calculation is needed in
order to determine the resulting wave form. This kind of geometry
could have interesting consequences for the solar neutrino signal,
because in this case the deviation from standard MSW analyses only
arises for neutrinos which travel sufficiently close to the solar
equatorial plane. Given the roughly 7-degree inclination of the
Earth's orbit relative to the plane of the solar equator, there is
a possibility of observing a seasonal dependence in the observed
solar neutrino flux. Since the presence of the MG resonance tends
to decrease the MSW effect, the prediction would be that the
observed rate of solar electron-neutrino events is maximized when
the Earth is closest to the solar equatorial plane (December and
June) and is minimized when furthest from this plane (March and
September).

\section*{Acknowledgements}

This work was supported by Spanish grants BFM2002-00345, by the
European Commission RTN network HPRN-CT-2000-00148, by the
European Science Foundation network grant N.~86, by Iberdrola
Foundation (VBS) and by INTAS grant YSF 2001/2-148 and CSIC-RAS
agreement (TIR). C.B.'s research is supported by grants from NSERC
(Canada), FCAR (Quebec) and McGill University. M.M.\ is supported
by contract HPMF-CT-2000-01008.  VBS, NSD and TIR were partially
supported by the RFBR grant 00-02-16271. C.B. would like to thank
the University of Valencia for its hospitality during part of this
work.

\end{document}